# Photoacoustics meets ultrasound: micro-Doppler photoacoustic effect and detection by ultrasound

Fei Gao[1,*], Xiaohua Feng[1], Yuanjin Zheng[1], and Claus-Dieter Ohl[2]

[1]School of Electrical and Electronic Engineering, Nanyang Technological University, 639798, Singapore

[2]School of Physical and Mathematical Sciences, Nanyang Technological University, 639798, Singapore

[*]Corresponding author: fgao1@e.ntu.edu.sg

**Abstract:**

In recent years, photoacoustics has attracted intensive research for both anatomical and functional biomedical imaging. However, the physical interaction between photoacoustic generated endogenous waves and an exogenous ultrasound wave is a largely unexplored area. Here, we report the initial results about the interaction of photoacoustic and external ultrasound waves leading to a micro-Doppler photoacoustic (mDPA) effect, which is experimentally observed and consistently modelled. It is based on a simultaneous excitation on the target with a pulsed laser and continuous wave (CW) ultrasound. The thermoelastically induced expansion will modulate the CW ultrasound and leads to transient Doppler frequency shift. The reported mDPA effect can be described as frequency modulation of the intense CW ultrasound carrier through photoacoustic vibrations. This technique may open the possibility to sensitively detect the photoacoustic vibration in deep optically and acoustically scattering medium, avoiding acoustic distortion that exists in state-of-the-art pulsed photoacoustic imaging systems.

Photoacoustic effect refers to the light-induced acoustic generation discovered by Alexander Bell in 1880 [1-2]. Based on photoacoustic effect, both photoacoustic microscopy (PAM) and photoacoustic tomography (PAT) have attracted increasing interest in recent years on multi-scale biomedical imaging research, ranging from molecular imaging of biomarkers to whole body imaging of small animals [3-10]. PAM and PAT circumvent the penetration depth limitation of conventional optical imaging modalities due to optical diffusion by listening to thermoelastically induced photoacoustic signals. In a typical photoacoustic imaging application, a nanosecond pulsed laser is employed to illuminate the biological tissue. Upon optical absorption and heating of endogenous chromophores (melanin, haemoglobin, etc.), transient acoustic waves are launched due to thermoelastic expansion. However, due to the low energy conversion efficiency of this process, acoustic attenuation, and scattering in heterogeneous biological tissues, the received acoustic signal is usually weak and severely distorted, limiting the endogenous contrast and resolution in photoacoustic imaging.

To enhance the signal-to-noise ratio (SNR), the pulse energy of laser may be increased, yet this is limited by the ANSI laser exposure standard for human safety (<20 mJ/cm$^2$ at 532 nm wavelength). The most commonly used approaches are pre-amplification of the signal and averaging after multiple data acquisition. The signal enhancement this way is still limited by long acquisition time, and the hardware's performance, e.g. limited gain and bandwidth and unavoidable instrument noise. Contrast-enhanced photoacoustic imaging is studied extensively with the help of exogenous contrast agents, such as nanoparticles [11-12], carbon nanotubes [13], and vaporized nanodroplets [14]. The problems encountered with these engineered contrast agents are their limited optical absorption and issues with toxicity. Besides increasing contrast, acoustic distortion in heterogeneous tissues is mostly addressed through algorithm re-design [15-17].

To potentially address the challenges of conventional pohotoacoustic imaging, here we report on a photoacoustic detection technique, which may increase the sensitivity without adding agents and avoid some of the distortions, by which photoacoustics is plagued. This is achieved by combining endogenous induced photoacoustic vibration and exogenous excited ultrasound, abbreviated as photoacoustic-ultrasound interaction. More specifically, a pulsed laser illumination and CW ultrasound driving are utilized simultaneously, it is observed that endogenous induced photoacousic vibration can modulate the exogenous excited ultrasound wave in terms of Doppler frequency shift. The detailed working principle is discussed as below.

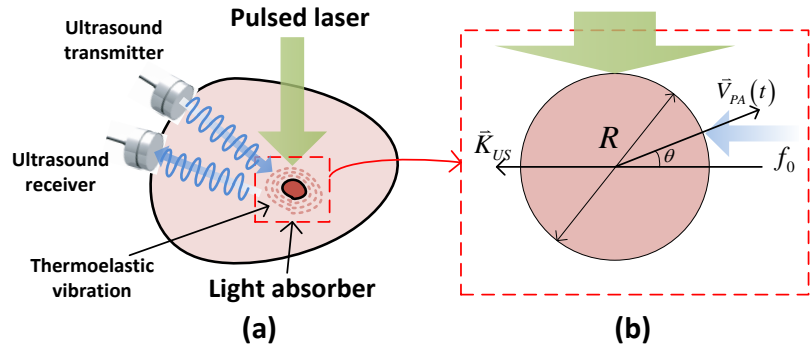

FIG. 1. The mDPA effect and modelling. (a) Fundamentals of photoacoustics-ultrasound interaction when light absorber is illuminated by both pulsed laser and CW ultrasound simultaneously. mDPA effect occurs when the laser-induced thermoelastic vibration modulates the CW ultrasound in terms of micro-Doppler frequency shift. (b) Simplified round-shape model of the light absorber with diameter $R$. Laser illumination is from top side, and ultrasound excitation is from right side.

Here we transmit the CW ultrasound and receive the backscattering ultrasound from an optical-absorbing object, which is excited with a pulsed laser to emit a photoacoustic pulse at the same time. Because the photoacoustic effect imparts a transient velocity change to the object due to thermoelastic expansion, the frequency of received CW ultrasound will be shifted due to Doppler effect. Being different from photoacoustic Doppler effect caused by bulk translation of the target at a constant velocity [18-20], the frequency shift induced by the mechanical vibration or rotation of the target is termed as micro-Doppler effect [21]. Interestingly, here we treat the photoacoustic effect as a thermoelastic mechanical vibration induced by pulsed laser illumination at the optical absorber site, rather than conventional photoacoustic wave directly detected by ultrasound transducer, see Fig. 1(a). Therefore, photoacoustic induced thermoelastic vibrations will modulate the frequency of the received CW ultrasound wave, which we define as micro-Doppler photoacoustic (mDPA) effect. As shown in Fig. 1(b), the model of the mDPA effect is simplified to be a plane CW ultrasound wave with frequency $f_0$ hitting on a round-shape target with diameter $R$. The transient velocity vector $\vec{V}_{PA}$ of the photoacoustic vibration is proportional to the derivative of the photoacoustic pressure $p(t)$ expressed as:

$$V_{PA}(t) = \kappa_s R \frac{\partial p(t)}{\partial t} \tag{1}$$

where $\kappa_s$ is the adiabatic compressibility. Then the transient micro-Doppler frequency shift can be expressed as:

$$f_{mDPA} = 2f_0 \frac{V_{PA}(t)}{c} \cos\theta = \frac{2f_0 \kappa_s R}{c} \frac{\partial p(t)}{\partial t} \cos\theta \tag{2}$$

where $\theta$ is the photoacoustic vibration angle annotated in Fig. 1(b). From Eq. (2) it is observed that the micro-Doppler frequency shift $f_{mDPA}$ is proportional to the derivative of the photoacoustic pressure $p(t)$, showing the feasibility of extracting the photoacoustic information from the micro-Doppler frequency shift of the ultrasound transmitter/receiver.

The proposed mDPA effect allows the weak and wideband pulsed photoacoustic signal to be carried by the intensive narrowband ultrasound wave, retaining much stronger immunity against acoustic attenuation and distortion in heterogeneous acoustic channel such as biological tissues.

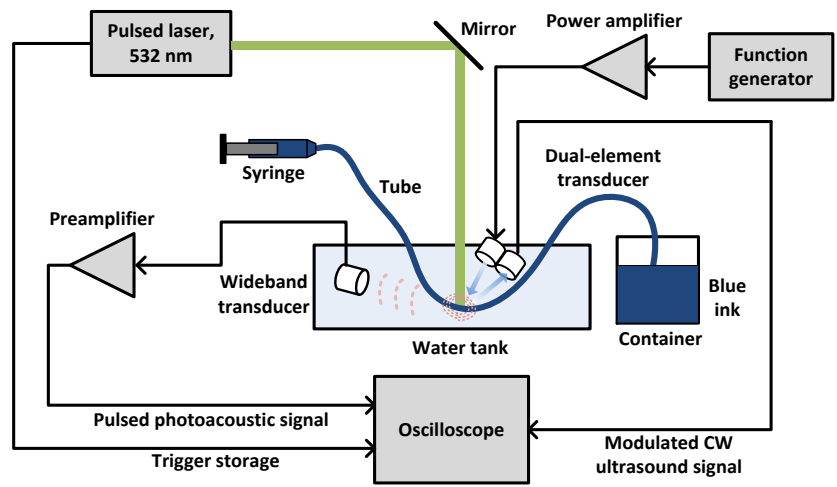

FIG. 2. Diagram of the experimental setup for the detection of conventional pulsed photoacoustic signal and CW ultrasound modulated by the micro-Doppler photoacoustic effect.

Next we will experimentally test if this Doppler shift can be picked up (Fig. 2). The setup consist of a Q-switched Nd: YAG laser at 532 nm emitting single laser pulses with 7ns pulse width (Orion, New Wave, Inc). The collimated laser beam with 2 mm diameter spot size is guided onto a silicone tube immersed in water filled with diluted blue ink (Pelikan, $\mu_a = \sim 100 \text{cm}^{-1}$) pumped with a syringe pump. Two US transceiver systems are used. First a wideband ultrasound transducer (V323-SU, 2.25 MHz, 6 mm in diameter; Olympus) is adopted to receive the photoacoustic wave conventionally and acts as a reference. It is connected to a preamplifier (54 dB gain, model 5662; Olympus) and the signal is recorded with an oscilloscope (500 MHz sampling rate, WaveMaster 8000A; LeCroy). The data is 20 times averaged. The second transceiver system is running simultaneously using a dual-element ultrasound transducer (5 MHz, 6 mm in diameter, DHC711-RM; Olympus). It operates as a CW ultrasound radar, transmitting and receiving CW ultrasound by its dual elements. A 5 MHz sine wave generated from a function generator is fed into a power amplifier (250 W; BT00200-AlphaSA-CW, Tomcorf) to drive the transmitting element. The receiving element is directly connected to the oscilloscope for data recording. Both the wideband and dual-element transducers are placed near the testing tube at a distance of 4.5 cm. A synchronization signal from the pulsed laser is used to trigger the data acquisition of the oscilloscope.

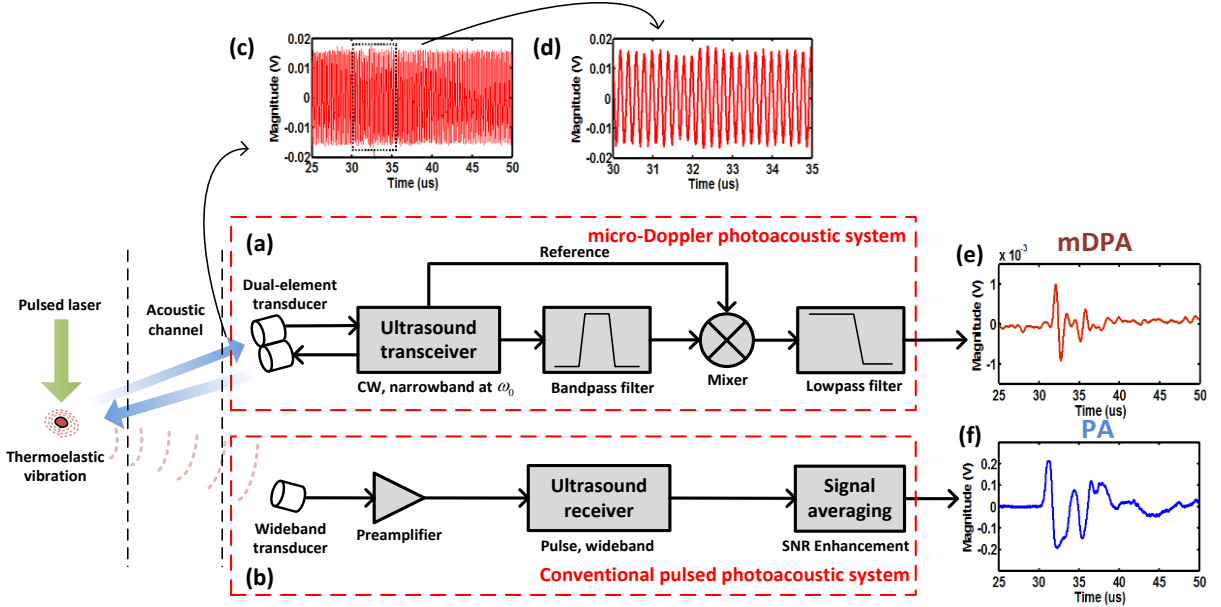

FIG. 3. The mDPA and pulsed photoacoustic system. (a) micro-Doppler photoacoustic system, and (b) conventional pulsed photoacoustic system. (c-d) Received CW ultrasound signal with frequency modulation by micro-Doppler photoacoustic effect, and (e) recovered micro-Doppler photoacoustic signal. (f) Conventional pulsed photoacoustic signal.

To extract the micro-Doppler frequency shift from the received ultrasound signal, band-pass filtering and down-conversion technique are employed, which multiplies the received ultrasound wave (Fig. 3(c-d)) $US(t) = A_{US} \sin\left[2\pi(f_0 + f_{mDPA})t\right]$ with a reference signal $R(t) = A_R \cos\left[2\pi f_0 t\right]$. Here $A_{US}$ and $A_R$ are the respective amplitudes. Then we apply low-pass filtering:

$$
\begin{aligned}
mDPA(t) = US(t)R(t) &= A_{US}A_R \sin\left[2\pi(f_0 + f_{mDPA})t\right]\cos\left[2\pi f_0 t\right] \\
&= \frac{1}{2}A_{US}A_R \sin\left[2\pi(2f_0 + f_{mDPA})t\right] + \frac{1}{2}A_{US}A_R \sin\left[2\pi f_{mDPA}t\right] \\
&\xrightarrow{Low-pass\ filtering} \frac{1}{2}A_{US}A_R \sin\left[2\pi f_{mDPA}t\right] \xrightarrow{Small-angle\ approximation} A_{US}A_R \pi f_{mDPA} t \\
&= \frac{2\pi f_0 \kappa_s R A_{US} A_R}{c}\cos\theta \frac{\partial p(t)}{\partial t} t = K \frac{\partial p(t)}{\partial t} t
\end{aligned}
\quad (3)
$$

where small-angle approximation is applicable due to the much shorter duration of photoacoustic wave than mDPA period, and $K = 2\pi f_0 \kappa_s R A_{US} A_R \cos\theta / c$ is assumed to be the mDPA conversion constant. Both photoacoustic (PA) wave (Fig. 3(f)) and mDPA wave (Fig. 3(e)) detected by conventional pulsed photoacoustic system (Fig. 3(b)) and the proposed mDPA system (Fig. 3(a)) respectively, indicate the photoacoustic source located 4.5 cm away with a delay of about 30 μs.

The extracted mDPA signal amplitude is proportional to the derivative of the photoacoustic pressure according to Eq. (3), so it is also expected to reveal the optical absorption coefficient as the conventional photoacoustic pressure does. The laser pulse energy is varied from 20 μJ to 90 μJ, to validate that the normalized amplitude of the mDPA wave has a good agreement with the amplitude of conventional photoacoustic wave for optical absorption measurement as shown in Fig. 4.

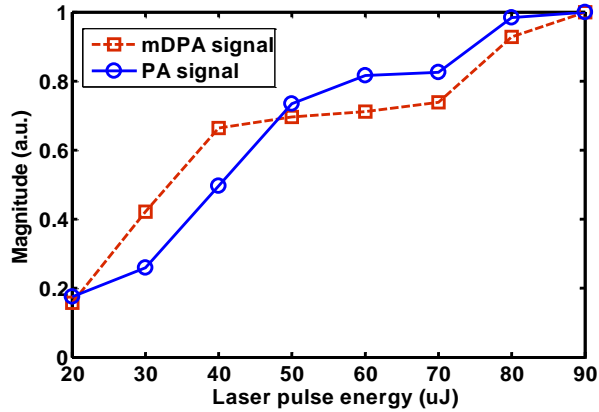

FIG. 4 Normalized amplitude comparison of micro-Doppler photoacoustic (mDPA) signal and pulsed photoacoustic (PA) signal by varying the laser pulse energy.

To compare the sensitivity of the proposed mDPA system with a conventional pulsed photoacoustic system, pre-amplification and averaging are removed from both the mDPA system and the conventional pulsed photoacoustic system. At the source site as shown on the left side of Fig. 5(a), both the mDPA signal and the pulsed photoacoustic signal are detectable above the noise floor. However, after acoustic attenuation and distortion during the propagation, the amplitude of the pulsed photoacoustic signal is below the input referred noise of ultrasound receiver, which is equivalently to the base-line noise floor of the oscilloscope used in the experiment as shown on the right of Fig. 5(a). Therefore, the pulsed photoacoustic wave can be hardly detected in presence of the background noise (Fig. 5(b)). On the other hand, the mDPA system is still capable of recovering the photoacoustic information (Fig. 5(c)), for the reason that photoacoustic vibrations modulate the intense CW ultrasound wave in terms of micro-Doppler frequency shift. The modulated CW ultrasound wave reserves the high intensity and the narrow bandwidth, which allows band-pass filtering with high quality factor (Q) to significantly suppress the noise. The experimental result shows higher signal SNR (14dB) and fidelity than pulsed photoacoustic signal SNR (2.5dB) during its propagation in acoustic channel.

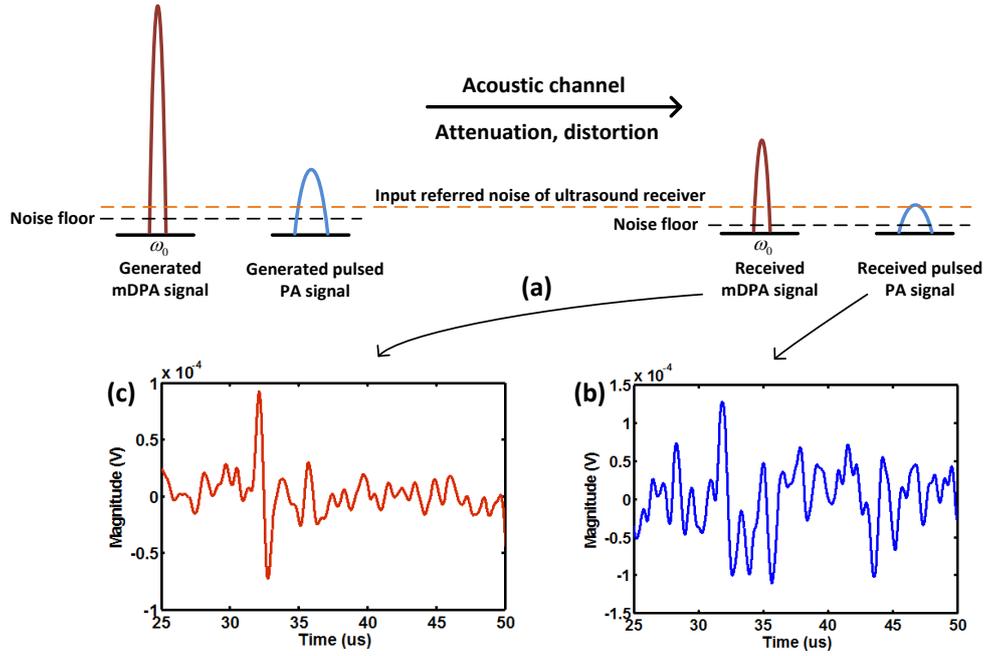

FIG. 5 SNR analysis of mDPA versus pulsed photoacoustic signals. (a) Signal and noise levels of micro-Doppler photoacoustic and pulsed photoacoustic signals before and after acoustic channel. (b) Waveform of received pulsed photoacoustic signal, and (c) recovered mDPA signal. The mDPA signal detection retains data... higher SNR than conventional pulsed photoacoustic detection.

To further illustrate the feature and advantage of the proposed mDPA system, it is interesting to make an analogy between the photoacoustic detection system and the well-established telecommunication system. More specifically, conventional pulsed photoacoustic detection is similar with amplitude modulation (AM) communication system, where information is carried in terms of signal's amplitude. On the other hand, the proposed mDPA detection can be related to a frequency modulation (FM) communication system, where information is carried in terms of signal's frequency shift. According to communication theory, FM systems are far better at rejecting noise as compared to AM system: the noise is distributed uniformly in frequency and varies randomly in amplitude. Therefore, the FM system is inherently immune to the random noise due to its narrowband characteristics and high Q band-pass filtering, guaranteeing the superior advantages of the FM mDPA system over conventional AM pulsed photoacoustic system.

In summary, the photoacoustics-ultrasound interaction is studied to experimentally observe the mDPA effect. Simultaneous illuminating the object by pulsed laser and CW ultrasound leads to the micro-Doppler frequency modulation of CW ultrasound induced by transient thermoelastic expansion and vibration. Secondly, based on the mDPA effect, a new photoacoustic detection system, termed micro-Doppler photoacoustic system, is proposed to extract the photoacoustic information from the received modulated CW ultrasound signal through down-conversion technique. Due to the mDPA effect, the weak and wideband photoacoustic signal is modulated onto the CW ultrasound carrier in terms of micro-Doppler frequency shift. Taking advantage of CW ultrasound's high intensity and narrowband spectrum, it retains much stronger immunity than the pulsed photoacoustic signal against the acoustic attenuation and distortion suffered during its propagation in the heterogeneous acoustic channel. Comparison studies demonstrate that mDPA signal could achieve much higher signal-to-noise (SNR) ratio and fidelity than conventional pulsed photoacoustic system. It is highly expected that an imaging system based on mDPA effect will outperform the conventional pulsed photoacoustic imaging, which will be studied in the future work.

This research is supported by the Singapore National Research Foundation under its Exploratory/Developmental Grant (NMRC/EDG/1062/2012) and administered by the Singapore Ministry of Health's National Medical Research Council.